\documentclass[%
reprint,
showpacs,preprintnumbers,
nofootinbib,
amsmath,amssymb,
aps,
prd,
floatfix,
]{revtex4-1}

\usepackage{graphicx}
\usepackage{dcolumn}
\usepackage{bm}

\usepackage{url}

\begin{document}

\preprint{\begin{tabular}{l}
\texttt{EURONU-WP6-10-29}
\\
\texttt{arXiv:1006.3244v3 [hep-ph]}
\end{tabular}}

\title{Statistical Significance of the Gallium Anomaly}

\author{Carlo Giunti}
\email{giunti@to.infn.it}
\altaffiliation[also at ]{Department of Theoretical Physics, University of Torino, Italy}
\affiliation{INFN, Sezione di Torino, Via P. Giuria 1, I--10125 Torino, Italy}

\author{Marco Laveder}
\email{laveder@pd.infn.it}
\affiliation{Dipartimento di Fisica ``G. Galilei'', Universit\`a di Padova,
and
INFN, Sezione di Padova,
Via F. Marzolo 8, I--35131 Padova, Italy}

\date{\today}

\begin{abstract}
We calculate the statistical significance of the anomalous deficit of
electron neutrinos measured in the radioactive source experiments of the
GALLEX and SAGE solar neutrino detectors
taking into account the uncertainty of the detection cross section.
We found that the statistical significance of the anomaly is about
$3.0\sigma$.
A fit of the data in terms of neutrino oscillations
favors at about $2.7\sigma$ short-baseline electron neutrino disappearance
with respect to the null hypothesis of no oscillations.
\end{abstract}

\pacs{14.60.Pq, 14.60.Lm, 14.60.St}

\maketitle

The GALLEX
\cite{Anselmann:1995ar,Hampel:1998fc,1001.2731}
and
SAGE
\cite{Abdurashitov:1996dp,hep-ph/9803418,nucl-ex/0512041,0901.2200}
Gallium solar neutrino experiments have been tested with
intense artificial ${}^{51}\text{Cr}$ and ${}^{37}\text{Ar}$ radioactive sources
placed inside the detectors.
The results of these
``Gallium radioactive source experiments''
indicate a ratio $R$ of measured and predicted ${}^{71}\text{Ge}$ event rates
which is smaller than unity:
\begin{align}
R^{\text{G1}}_{\text{B}}
=
\null & \null
0.953 \pm 0.11
\,,
\label{001}
\\
R^{\text{G2}}_{\text{B}}
=
\null & \null
0.812 {}^{+0.10}_{-0.11}
\,,
\label{002}
\\
R^{\text{S1}}_{\text{B}}
=
\null & \null
0.95 \pm 0.12
\,,
\label{003}
\\
R^{\text{S2}}_{\text{B}}
=
\null & \null
0.791 {}^{+0.084}_{-0.078}
\,,
\label{004}
\end{align}
where
$\text{G1}$ and $\text{G2}$ denote the two GALLEX experiments with ${}^{51}\text{Cr}$ sources,
$\text{S1}$ denotes the SAGE experiment with a ${}^{51}\text{Cr}$ source,
and
$\text{S2}$ denotes the SAGE experiment with a ${}^{37}\text{Ar}$ source.

\begin{figure}[t!]
\begin{center}
\includegraphics*[bb=8 14 571 565, width=\linewidth]{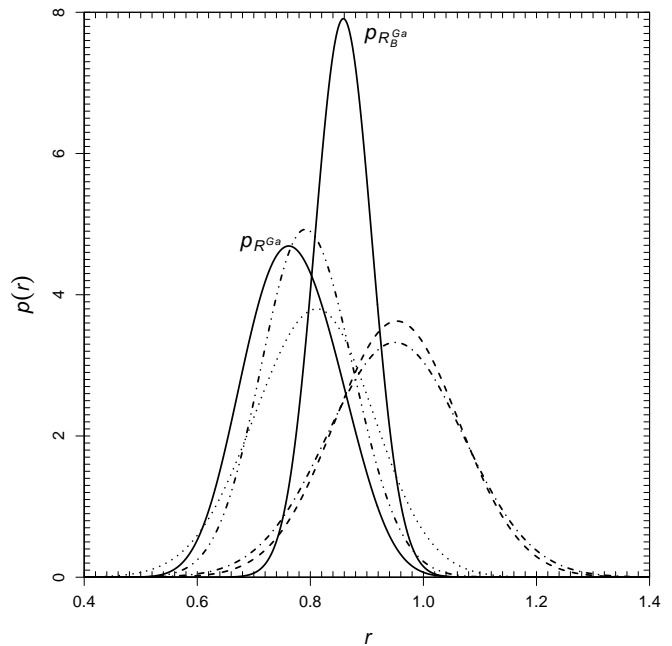}
\end{center}
\caption{ \label{019}
Solid lines: probability distributions $p_{R^{\text{Ga}}_{\text{B}}}(r)$ and $p_{R^{\text{Ga}}}(r)$,
as indicated by the labels.
Dashed, dotted, dash-dotted and dash-dot-dotted lines: probability distributions
$p_{R^{\text{G1}}_{\text{B}}}(r)$,
$p_{R^{\text{G2}}_{\text{B}}}(r)$,
$p_{R^{\text{S1}}_{\text{B}}}(r)$ and
$p_{R^{\text{S2}}_{\text{B}}}(r)$, respectively.
}
\end{figure}

Assuming Gaussian probability distributions and
taking into account the asymmetric uncertainties of
$R^{\text{G2}}_{\text{B}}$
and
$R^{\text{S2}}_{\text{B}}$,
we have the probability distributions shown by the
dashed, dotted, dash-dotted and dash-dot-dotted lines in Fig.~\ref{019}.
The combined probability distribution $p_{R^{\text{Ga}}_{\text{B}}}(r)$
shown in Fig.~\ref{019} gives the average ratio
\begin{equation}
R^{\text{Ga}}_{\text{B}}
=
0.86
{}^{+0.05}_{-0.05}
{}^{+0.10}_{-0.10}
{}^{+0.15}_{-0.15}
\,,
\label{005}
\end{equation}
where the uncertainties are at
68.27\% C.L. ($1\sigma$),
95.45\% C.L. ($2\sigma$),
99.73\% C.L. ($3\sigma$).
Thus,
the number of measured events is about $2.8\sigma$ smaller than the prediction.
This is the ``Gallium anomaly''.

As indicated by the ``B'' subscript,
the ratios in Eqs.~(\ref{001})--(\ref{005}) have been calculated with respect to the
rate estimated using the best-fit values of the cross section of the detection process
\begin{equation}
\nu_{e} + {}^{71}\text{Ga} \to {}^{71}\text{Ge} + e^{-}
\label{006}
\end{equation}
calculated by Bahcall \cite{hep-ph/9710491},
\begin{align}
\sigma_{\text{B}}^{\text{bf}}({}^{51}\text{Cr})
=
\null & \null
\left( 58.1 {}^{+2.1}_{-1.6} \right) \times 10^{-46} \, \text{cm}^2
\,,
\label{007}
\\
\sigma_{\text{B}}^{\text{bf}}({}^{37}\text{Ar})
=
\null & \null
\left( 70.0 {}^{+4.9}_{-2.1} \right) \times 10^{-46} \, \text{cm}^2
\,.
\label{008}
\end{align}
The uncertainties of these cross sections are not taken into account in the experimental ratios
in Eqs.~(\ref{001})--(\ref{004}).
These uncertainties are large
\cite{nucl-th/9503017,hep-ph/9710491,nucl-th/9804011},
because only the cross section of the transition
from the ground state of ${}^{71}\text{Ga}$ to the ground state of ${}^{71}\text{Ge}$
is known with precision from the measured rate of electron capture decay of
${}^{71}\text{Ge}$ to ${}^{71}\text{Ga}$.
Electron neutrinos produced by ${}^{51}\text{Cr}$ and ${}^{37}\text{Ar}$ radioactive sources
can be absorbed also through transitions from the ground state of ${}^{71}\text{Ga}$
to two excited states of ${}^{71}\text{Ge}$
at 175 keV and 500 keV,
with cross sections which are inferred using a nuclear model
from $ p + {}^{71}\text{Ga} \to {}^{71}\text{Ge} + n $ measurements
\cite{Krofcheck:1985fg}.

Hence, at least part of the deficit of measured events with respect to the prediction
could be explained by an overestimation of the transitions
to the two excited states of ${}^{71}\text{Ge}$
\cite{nucl-ex/0512041,hep-ph/0605186,0901.2200}.
However,
since the contribution of the transitions to the two excited states of ${}^{71}\text{Ge}$ is only 5\%
\cite{hep-ph/9710491},
even the complete absence of such transitions
would reduce the ratio of measured and predicted ${}^{71}\text{Ge}$ event rates to about
$0.91\pm0.05$,
leaving an anomaly of about $1.7\sigma$
\cite{1005.4599}.

We think that for a correct assessment of the statistical significance of the Gallium anomaly
simple approaches based on either accepting the Bahcall cross section in Eq.~(\ref{007})
without taking into account its uncertainty or
suppressing without theoretical motivations the transitions to the two excited states of ${}^{71}\text{Ge}$
are insufficient.
A correct assessment of the statistical significance of the Gallium anomaly
can be done by taking into account the large uncertainties of the transitions to the two excited states of ${}^{71}\text{Ge}$
\cite{nucl-th/9503017,hep-ph/9710491,nucl-th/9804011}.
The most reliable estimate of these transitions and their uncertainties have been done by Haxton in Ref.~\cite{nucl-th/9804011},
leading to the total cross section for a ${}^{51}\text{Cr}$ source
\begin{equation}
\sigma_{\text{H}}({}^{51}\text{Cr}) = \left( 63.9 \pm 6.8 \right) \times 10^{-46} \, \text{cm}^2
\,.
\label{009}
\end{equation}
Notice that the average value of this cross section is even larger than the Bahcall
cross section in Eq.~(\ref{007}).
This leads to an enhancement of the Gallium anomaly.
However, the uncertainty of $\sigma_{\text{H}}({}^{51}\text{Cr})$ is rather large.
Hence, a correct assessment of the statistical significance of the Gallium anomaly
requires an accurate treatment of the cross section uncertainty.

Since the ratios in Eqs.~(\ref{001})--(\ref{003}) have been calculated with respect to the
best-fit value in Eq.~(\ref{007}) of the Bahcall cross section for a ${}^{51}\text{Cr}$ source,
these ratios must be rescaled by
\begin{equation}
R^{\text{H}}_{\text{B}}({}^{51}\text{Cr})
=
\frac{\sigma_{\text{H}}({}^{51}\text{Cr})}{\sigma_{\text{B}}^{\text{bf}}({}^{51}\text{Cr})}
=
1.10 \pm 0.12
\,.
\label{010}
\end{equation}

For the SAGE ${}^{37}\text{Ar}$ source experiment,
we evaluate the detection cross section and its uncertainty as follows.
The cross section is given by
\cite{hep-ph/9710491}
\begin{equation}
\sigma({}^{37}\text{Ar})
=
\sigma_{\text{gs}}
\left(
1
+
0.695
\frac{\text{BGT}_{175}}{\text{BGT}_{\text{gs}}}
+
0.263
\frac{\text{BGT}_{500}}{\text{BGT}_{\text{gs}}}
\right)
\,,
\label{011}
\end{equation}
where
$\sigma_{\text{gs}}=66.2\times10^{-46}\,\text{cm}^2$
is the cross section from the ground state of ${}^{71}\text{Ga}$ to the ground state of ${}^{71}\text{Ge}$,
$\text{BGT}_{\text{gs}}$
is the corresponding Gamow-Teller strength
and
$\text{BGT}_{175}$
and
$\text{BGT}_{500}$
are the Gamow-Teller strengths of the transitions
from the ground state of ${}^{71}\text{Ga}$ to the two excited states of ${}^{71}\text{Ge}$
at 175 keV and 500 keV.
The coefficients of
$\text{BGT}_{175}/\text{BGT}_{\text{gs}}$
and
$\text{BGT}_{500}/\text{BGT}_{\text{gs}}$
are determined by phase space.
In Ref.~\cite{nucl-th/9804011}, Haxton estimated\footnote{
In Ref.~\cite{nucl-th/9804011},
the values of
$\text{BGT}_{175}/\text{BGT}_{\text{gs}}$
and
$\text{BGT}_{500}/\text{BGT}_{\text{gs}}$
can be extracted, respectively, from Eqs.~(12) and (7),
taking into account Eq.~(1).
As explained by Haxton,
$\text{BGT}_{175}$ requires a theoretical calculation,
whereas
for $\text{BGT}_{500}$ it is reasonable to adopt the corresponding $(p,n)$ value.
}
\begin{align}
\null & \null
\text{BGT}_{175}/\text{BGT}_{\text{gs}} = 0.19 \pm 0.18
\,,
\label{012}
\\
\null & \null
\text{BGT}_{500}/\text{BGT}_{\text{gs}} = 0.13 \pm 0.02
\,.
\label{013}
\end{align}
Thus, we obtain
\begin{equation}
\sigma_{\text{H}}({}^{37}\text{Ar})
=
\left( 77.3 \pm 8.2 \right) \times 10^{-46} \, \text{cm}^2
\,,
\label{014}
\end{equation}
and
\begin{equation}
R^{\text{H}}_{\text{B}}({}^{37}\text{Ar})
=
\frac{\sigma_{\text{H}}({}^{37}\text{Ar})}{\sigma_{\text{B}}^{\text{bf}}({}^{37}\text{Ar})}
=
1.10 \pm 0.12
\,,
\label{015}
\end{equation}
which has the same value of $R^{\text{H}}_{\text{B}}({}^{51}\text{Cr})$ in Eq.~(\ref{010}).
Therefore,
all the ratios in Eqs.~(\ref{001})--(\ref{004}) must be rescaled by
$
R^{\text{H}}_{\text{B}}
=
R^{\text{H}}_{\text{B}}({}^{51}\text{Cr})
=
R^{\text{H}}_{\text{B}}({}^{37}\text{Ar})
$.

One must also take into account that
the value of the cross section is bounded from below by the
cross section $\sigma_{\text{gs}}$ of the transition
from the ground state of ${}^{71}\text{Ga}$ to the ground state of ${}^{71}\text{Ge}$
\cite{hep-ph/9710491}:
\begin{equation}
R^{\text{H}}_{\text{B}}
\geq
R^{\text{gs}}_{\text{B}}
=
\frac{\sigma_{\text{gs}}}{\sigma_{\text{B}}^{\text{bf}}}
=
0.95
\,.
\label{016}
\end{equation}

In the following we calculate the probability distribution of
\begin{equation}
R^{\text{Ga}}
=
\frac{R^{\text{Ga}}_{\text{B}}}{R^{\text{H}}_{\text{B}}}
\label{017}
\end{equation}
by taking into account the uncertainty of the denominator
$R^{\text{H}}_{\text{B}}$
given in Eqs.~(\ref{010}) and (\ref{015}).
This is the theoretical uncertainty of the cross section
that has not been taken into account in the ratios
(\ref{001})--(\ref{005}),
which have been evaluated using the best-fit values of the Bahcall cross sections in Eqs.~(\ref{007}) and (\ref{008}).
We assume a Gaussian probability distribution truncated below $R^{\text{gs}}_{\text{B}}$:
\begin{equation}
p_{R^{\text{H}}_{\text{B}}}(r)
\propto
\left\{
\begin{array}{ll} \displaystyle
\exp\!\left[
- \frac{1}{2}
\left(
\frac{r-\langle R^{\text{H}}_{\text{B}} \rangle}{\Delta R^{\text{H}}_{\text{B}}}
\right)^2
\right]
\,,
&
r \geq R^{\text{gs}}_{\text{B}}
\,,
\\ \displaystyle
0
\,,
&
r < R^{\text{gs}}_{\text{B}}
\,,
\end{array}
\right.
\label{018}
\end{equation}
with
$\langle R^{\text{H}}_{\text{B}} \rangle = 1.10$
and
$\Delta R^{\text{H}}_{\text{B}} = 0.12$.

The probability distribution of the ratio $R^{\text{Ga}}$ in Eq.~(\ref{017}) is given by
(see Section~2.4.4 of Ref.~\cite{James:2006zz})
\begin{equation}
p_{R^{\text{Ga}}}(r)
=
\int_{R^{\text{gs}}_{\text{B}}}^{\infty}
p_{R^{\text{Ga}}_{\text{B}}}(rs)
\,
p_{R^{\text{H}}_{\text{B}}}(s)
\,
s
\,
\text{d}s
\,.
\label{020}
\end{equation}
Figure~\ref{019} shows the probability distribution
$p_{R^{\text{Ga}}_{\text{B}}}(r)$
of $R^{\text{Ga}}_{\text{B}}$ derived from the experimental data in Eqs.~(\ref{001})--(\ref{004})
and the result of the integral in Eq.~(\ref{020}).
One can see that $p_{R^{\text{Ga}}}(r)$ is peaked at a smaller value than $p_{R^{\text{Ga}}_{\text{B}}}(r)$,
but the uncertainty is larger.
We obtain
\begin{equation}
R^{\text{Ga}}
=
0.76
{}^{+0.09}_{-0.08}
{}^{+0.17}_{-0.15}
{}^{+0.24}_{-0.21}
\,,
\label{021}
\end{equation}
where the uncertainties are at
68.27\% C.L. ($1\sigma$),
95.45\% C.L. ($2\sigma$),
99.73\% C.L. ($3\sigma$).
From a comparison of these uncertainties
and from Fig.~\ref{019} one can see that the probability distribution is approximately Gaussian,
with slightly asymmetric uncertainties and tails which decrease slightly faster than Gaussian tails.

The probability of $R^{\text{Ga}}<1$ is $99.86\%$
($3.0\sigma$ anomaly),
slightly larger than the probability of $R^{\text{Ga}}_{\text{B}}<1$,
which is $99.75\%$
($2.8\sigma$ anomaly).
Therefore, the Gallium anomaly remains statistically significant after taking properly into account the
cross section uncertainty.

For the four individual Gallium radioactive source experiments,
using the same method as above,
from the experimental values in Eqs.~(\ref{001})--(\ref{004})
we obtain
\begin{align}
R^{\text{G1}}
=
R^{\text{G1}}_{\text{B}} / R^{\text{H}}_{\text{B}}
=
\null & \null
0.84
{}^{+0.13}_{-0.12}
{}^{+0.26}_{-0.23}
{}^{+0.40}_{-0.33}
\,,
\label{022}
\\
R^{\text{G2}}
=
R^{\text{G2}}_{\text{B}} / R^{\text{H}}_{\text{B}}
=
\null & \null
0.71
{}^{+0.12}_{-0.11}
{}^{+0.24}_{-0.21}
{}^{+0.36}_{-0.31}
\,,
\label{023}
\\
R^{\text{S1}}
=
R^{\text{S1}}_{\text{B}} / R^{\text{H}}_{\text{B}}
=
\null & \null
0.84
{}^{+0.14}_{-0.13}
{}^{+0.28}_{-0.24}
{}^{+0.42}_{-0.35}
\,,
\label{024}
\\
R^{\text{S2}}
=
R^{\text{S2}}_{\text{B}} / R^{\text{H}}_{\text{B}}
=
\null & \null
0.70
{}^{+0.10}_{-0.09}
{}^{+0.21}_{-0.17}
{}^{+0.31}_{-0.25}
\,,
\label{025}
\end{align}
with
$1\sigma$,
$2\sigma$,
$3\sigma$
uncertainties.
A comparison of these uncertainties
shows that the probability distributions are approximately Gaussian,
with slightly asymmetric uncertainties.

Since the Gallium anomaly is confirmed by the new statistical analysis which takes into account
the uncertainty of the detection cross section,
it is plausible that it is due to a physical mechanism.
In the following, we consider the possibility of electron neutrino disappearance
due to short-baseline oscillations
\cite{hep-ph/9411414,Laveder:2007zz,hep-ph/0610352,0707.4593,0711.4222,0902.1992,1005.4599,1006.2103}
(another explanation based on quantum decoherence in neutrino oscillations
has been proposed in Ref.~\cite{0805.2098}).

We consider the electron neutrino survival probability
\begin{equation}
P_{\nu_{e}\to\nu_{e}}^{\text{SBL}}(L,E)
=
1
-
\sin^2 2\vartheta
\sin^2\!\left( \frac{ \Delta{m}^2 L }{ 4 E } \right)
\,,
\label{026}
\end{equation}
where
$\vartheta$ is the mixing angle,
$\Delta{m}^2$ is the squared-mass difference,
$L$ is the neutrino path length and $E$ is the neutrino energy.
This survival probability is effective in short-baseline (SBL) experiments
in the framework of four-neutrino mixing schemes
(see Refs.~\cite{hep-ph/9812360,hep-ph/0405172,hep-ph/0606054,GonzalezGarcia:2007ib}),
which are the simplest extensions of three-neutrino mixing schemes which can accommodate
the two measured small solar and atmospheric squared-mass differences
$
\Delta{m}^2_{\text{SOL}}
\simeq
8 \times 10^{-5} \, \text{eV}^2
$
and
$
\Delta{m}^2_{\text{ATM}}
\simeq
2 \times 10^{-3} \, \text{eV}^2
$
and one larger squared-mass difference for short-baseline neutrino oscillations,
$
\Delta{m}^2 \gtrsim 0.1 \, \text{eV}^2
$.
The existence of a fourth massive neutrino corresponds,
in the flavor basis,
to the existence of a sterile neutrino $\nu_{s}$.

We performed a maximum likelihood analysis
(see Ref.~\cite{PDG-2008})
of the Gallium data as follows\footnote{
A standard least-squares analysis
would lead to misleading results,
because it does not allow us to take into account the lower bound
in Eq.~(\ref{016})
for $R^{\text{H}}_{\text{B}}$.
}.
We started with the calculation,
for each experiment,
of the value of
the ratio of the event rate as a function of
$\sin^2 2\vartheta$
and
$\Delta{m}^2$
and the event rate in absence of neutrino oscillations
(see Ref.~\cite{0711.4222} for details):
\begin{equation}
R^{k}(\sin^2 2\vartheta, \Delta{m}^2)
=
\dfrac
{ \int_{k} \text{d}V \, L^{-2} \sum_{i} b^{k}_{i} \, \sigma^{k}_{i} \, P_{\nu_{e}\to\nu_{e}}^{\text{SBL}}(L,E_{i}) }
{ \sum_{i} b^{k}_{i} \, \sigma^{k}_{i} \int_{k} \text{d}V \, L^{-2} }
\,,
\label{ratio}
\end{equation}
where the index $k$ labels the experiments
($
k
=
\text{G1},
\text{G2},
\text{S1},
\text{S2}
$),
the index $i$ labels the $\nu_{e}$ lines emitted in
${}^{51}\text{Cr}$ or ${}^{37}\text{Ar}$ electron captures
with energies $E_{i}$,
$b^{k}_{i}$ and $\sigma^{k}_{i}$ are the corresponding
branching ratios and cross sections
(see Table~I of Ref.~\cite{0711.4222}),
$L$ is the neutrino path length
and $\int_{k} \text{d}V$ is the integral over the volume of each detector
(see Table~II of Ref.~\cite{0711.4222}).

The uncertainty of $R^{\text{H}}_{\text{B}}$
is correlated in the calculation of the combined probability distribution of the four experimental ratios
in Eqs.~(\ref{022})--(\ref{025}).
Using a method similar to that utilized for the derivation of Eq.~(\ref{020})
(see Section~2.4.4 of Ref.~\cite{James:2006zz}),
we obtain the combined probability distribution
\begin{equation}
p_{\vec{R}}(\vec{r})
=
\int_{R^{\text{gs}}_{\text{B}}}^{\infty}
\left[
\prod_{k}
p_{R^{k}_{\text{B}}}(r^{k} s)
\right]
\,
p_{R^{\text{H}}_{\text{B}}}(s)
\,
s^4
\,
\text{d}s
\,,
\label{combined-probability}
\end{equation}
where
$\vec{R} = (R^{\text{G1}}, R^{\text{G2}}, R^{\text{S1}}, R^{\text{S2}})$
and
$\vec{r} = (r^{\text{G1}}, r^{\text{G2}}, r^{\text{S1}}, r^{\text{S2}})$.
The authors of Ref.~\cite{1101.2755}
considered a correlation of the systematic errors of the two GALLEX experiments and the two SAGE experiments.
Since such correlation is not documented in the
experimental publications,
where the combined ratio was calculated as a weighted average, without correlations,
we adopt the conservative approach of considering the systematic experimental errors as independent\footnote{
A correlation of the systematic experimental errors
can be taken into account in Eq.~(\ref{combined-probability})
by replacing
$
\prod_{k}
p_{R^{k}_{\text{B}}}(r^{k} s)
$
with a multivariate Gaussian distribution with the appropriate covariance matrix.
}.

\begin{figure}[t!]
\begin{center}
\includegraphics*[bb=5 11 571 571, width=\linewidth]{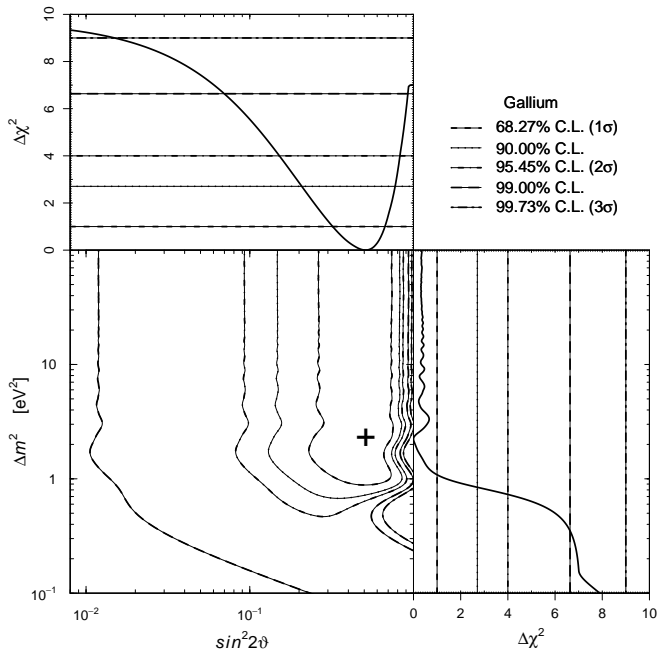}
\end{center}
\caption{ \label{027}
Allowed regions in the
$\sin^{2}2\vartheta$--$\Delta{m}^{2}$ plane
and
marginal $\Delta\chi^{2}$'s
for
$\sin^{2}2\vartheta$ and $\Delta{m}^{2}$
obtained from the
combined fit of the results of
the two GALLEX ${}^{51}\text{Cr}$ radioactive source experiments
and
the SAGE
${}^{51}\text{Cr}$ and ${}^{37}\text{Ar}$ radioactive source experiments.
The best-fit point corresponding to $\chi^2_{\text{min}}$ is indicated by a cross.
}
\end{figure}

The likelihood function of the
oscillation parameters
$\sin^2 2\vartheta$ and $\Delta{m}^2$
is given by
\begin{equation}
\mathcal{L}(\sin^{2}2\vartheta, \Delta{m}^{2})
=
p_{\vec{R}}(\vec{R}(\sin^2 2\vartheta, \Delta{m}^2))
\,,
\label{likelihood}
\end{equation}
with the four components of
$\vec{R}(\sin^2 2\vartheta, \Delta{m}^2)$
given by Eq.~(\ref{ratio}).
Figure~\ref{027} shows the allowed regions
in the
$\sin^{2}2\vartheta$--$\Delta{m}^{2}$ plane
and
the
marginal $\Delta\chi^{2} = \chi^2 - \chi^2_{\text{min}}$'s
for
$\sin^{2}2\vartheta$ and $\Delta{m}^{2}$,
from which one can infer the corresponding uncorrelated allowed intervals.
In the maximum likelihood analysis $\chi^{2}(\sin^{2}2\vartheta, \Delta{m}^{2})$
is given by
$-2\ln\mathcal{L}(\sin^{2}2\vartheta, \Delta{m}^{2})+\text{constant}$.

The best-fit values of the oscillation parameters are
\begin{equation}
\sin^2 2\vartheta_{\text{bf}} = 0.50
\,,
\quad
\Delta{m}^2_{\text{bf}} = 2.24 \, \text{eV}^2
\,.
\label{028}
\end{equation}
The value of the likelihood ratio between the null hypothesis of no oscillations and the oscillation hypothesis,
\begin{equation}
\frac{\mathcal{L}_{0}}{\mathcal{L}(\sin^2 2\vartheta_{\text{bf}},\Delta{m}^2_{\text{bf}})}
=
8
\times
10^{-3}
\,,
\label{029}
\end{equation}
is in favor of the oscillation hypothesis.
It corresponds to
$\Delta\chi^2 = 9.7$,
which, with two degrees of freedom,
disfavors the null hypothesis of no oscillations at
99.23\% C.L.
($2.7\sigma$).
The small difference between this statistical significance
of the indication in favor of the Gallium anomaly
and that obtained from Eq.~(\ref{021})
($3.0\sigma$)
is due to the different analysis of the data.
Although the neutrino oscillation analysis
leads to a better fit of the four data in Eqs.~(\ref{022})--(\ref{025})
(the best-fit values of the oscillation parameters in Eq.~(\ref{028}) give
$R^{\text{G1}}=0.75$,
$R^{\text{G2}}=0.75$,
$R^{\text{S1}}=0.73$ and
$R^{\text{S2}}=0.72$),
the correlation of the theoretical uncertainty of $R^{\text{H}}_{\text{B}}$
slightly disfavors a fit in which the deviations of the data from the best-fit values
do not have the same sign.

From Fig.~\ref{027} one can see that the marginal distributions of
$\sin^{2}2\vartheta$ and $\Delta{m}^{2}$
indicate that\footnote{
These bounds are weaker than those presented in a previous version of this paper
(\texttt{arXiv:1006.3244v2})
in which the correlation of the uncertainty of $R^{\text{H}}_{\text{B}}$
in the calculation of the combined probability distribution of the four experimental ratios
in Eqs.~(\ref{022})--(\ref{025})
was not taken into account.
}
\begin{equation}
\sin^{2}2\vartheta > 0.07
\,,
\quad
\Delta{m}^{2} > 0.35 \, \text{eV}^2
\,,
\label{030}
\end{equation}
at 99\% C.L..
These bounds indicate that the short-baseline disappearance of electron neutrinos
may be larger than that of electron antineutrinos,
which is bounded by the results of reactor neutrino experiments
\cite{0711.4222,0902.1992,1005.4599}.
This could be an indication of a violation of the CPT symmetry
\cite{1008.4750}
(CPT implies that $P_{\nu_{\alpha}\to\nu_{\alpha}}=P_{\bar\nu_{\alpha}\to\bar\nu_{\alpha}}$
for any flavor $\alpha=e,\mu,\tau$;
see Ref.~\cite{Giunti-Kim-2007}).
However,
according to a recent calculation
\cite{1101.2663}
the $\bar\nu_{e}$ fluxes produced in nuclear reactors are about 3\% larger than the standard ones
used in the analysis of reactor antineutrino data
(see Ref.~\cite{hep-ph/0107277}).
A comparison of the new reactor antineutrino fluxes with the data of several reactor neutrino experiments
suggests the existence of a reactor antineutrino anomaly \cite{1101.2755}
which is compatible with the Gallium anomaly
in a standard CPT-invariant neutrino oscillation framework.
In this case, the indication in favor of CPT violation obtained by comparing the
results of the neutrino oscillation analysis of Gallium and reactor data
is weakened,
but the plausibility of the existence of a Gallium anomaly is reinforced
by its compatibility with the reactor antineutrino anomaly.

CPT violation in short-baseline electron neutrino disappearance
can be tested with high accuracy in future experiments
with pure and well-known $\nu_{e}$ and $\bar\nu_{e}$ beams,
as beta-beam \cite{0907.3145}
and
neutrino factory \cite{0907.5487,1005.3146}
experiments.
Although the possibility of CPT violation is theoretically problematic
\cite{hep-ph/0309309},
it cannot be dismissed in phenomenological analyses of experimental results.
It is interesting to notice that
recently another indication of a violation of the CPT symmetry
has been found in the MINOS long-baseline $\nu_{\mu}$ and $\bar\nu_{\mu}$
disappearance experiment \cite{0910.3439,MINOS-Neutrino2010}.

There is also a growing experimental interest in favor of possible tests of the
Gallium anomaly.
In addition to the future experimental possibilities to test the short-baseline disappearance of electron neutrinos
discussed in Ref.~\cite{1005.4599},
the authors of Ref.~\cite{1006.2103} presented recently a plan to make an improved direct measurement of the Gallium anomaly
with the liquid Gallium metal used in the SAGE experiment and a new vessel divided in two zones,
which can measure a variation of the electron neutrino disappearance with distance.
The Borexino collaboration is studying the possibility of a
radioactive source experiment
\cite{hep-ex/9901012}
which could provide a ``smoking gun'' signal
by measuring the oscillation pattern inside the detector.
Other possible measurements with radioactive sources and different detector types
has been recently discussed in Ref.~\cite{1011.4509}.

The existence of at least four massive neutrinos,
one of which has a mass larger than about
$ 0.6 \, \text{eV} $
in order to generate
the squared-mass difference in Eq.~(\ref{030}),
can have important implications for cosmology
(see Refs.~\cite{hep-ph/0202122,astro-ph/0603494,0809.0631}).
The current indications of cosmological data analyzed in the framework of the
standard cosmological model
are controversial.
On one hand,
there are indications that the effective number of neutrino species may be larger than three
from Big Bang Nucleosynthesis \cite{1001.4440}
and from the Cosmic Microwave Background Radiation \cite{1001.4538}.
This is consistent with a thermalization of sterile neutrinos due to
active-sterile oscillations before Big Bang Nucleosynthesis
induced by the large values of the mixing parameters in Eq.~(\ref{030}) \cite{hep-ph/0308083}.
On the other hand,
analyses of Cosmic Microwave Background Radiation data and Large Scale Structure data
constrain the mass of a fourth thermalized neutrino to be smaller than about 0.7 eV
\cite{astro-ph/0511500,astro-ph/0607101,1006.5276,1102.4774}.
Hence,
either the heavy neutrino mass is close to the standard cosmological bound
or
the existence of short-baseline neutrino oscillations
is connected with non-standard
cosmological effects, as those discussed in Refs.~\cite{0812.2249,0812.4552,0906.3322}.

In conclusion,
we have estimated the uncertainty of the deficit of
electron neutrinos measured in the radioactive source experiments of the
GALLEX
\cite{Anselmann:1995ar,Hampel:1998fc,1001.2731}
and
SAGE
\cite{Abdurashitov:1996dp,hep-ph/9803418,nucl-ex/0512041,0901.2200}
solar neutrino detectors
taking into account the uncertainty of the detection cross section
estimated by Haxton in Ref.~\cite{nucl-th/9804011}.
The result shows that the Gallium anomaly is statistically significant,
at a level of about
$3.0\sigma$.
The analysis of the data in terms of neutrino oscillations
indicates values of the oscillation amplitude
$\sin^{2}2\vartheta \gtrsim 0.07$
and
squared-mass difference
$\Delta{m}^{2} \gtrsim 0.35 \, \text{eV}^2$
at 99\% C.L..

\bigskip
\centerline{\textbf{Acknowledgments}}
\medskip

We would like to thank
E.~Bellotti,
S.M.~Bilenky,
A.~Ianni,
T.~Lasserre,
E.~Lisi,
A.~Melchiorri,
G.~Ranucci,
S.~Schoenert,
T.~Schwetz,
and
C.~Volpe
for interesting discussions.

\bibliography{bibtex/nu,gal}

\end{document}